\documentclass[aps,pra,twocolumn,groupedaddress,showpacs]{revtex4}
\usepackage{dcolumn}  
\usepackage{amsmath}
\usepackage{bm}
\usepackage{CJK}      
\usepackage{ulem}     
\usepackage{color}    
\usepackage{makeidx}
\makeindex
\begin{document}
\begin{CJK*}{UTF8}{bsmi}

\preprint{Version 10.0}

\title{Simplified partial wave expansion of the Lamb shift}

\author{J. Sapirstein}
\email[]{jsapirst@nd.edu}
\affiliation{Department of Physics and Astronomy,
University of Notre Dame, Notre Dame, IN 46556-5670}

\author{K. T. Cheng (鄭國錚)}
\affiliation{Lawrence Livermore National Laboratory, Livermore, CA 94550}

\begin{abstract}

A method for calculating the self energy part of the Lamb shift is revisited.
When the electron propagator in an external field is represented as an expansion
in partial waves, the original method converges relatively slowly,
requiring the calculation of dozens of partial waves.
Here we show an improved method in which accurate results can be obtained
using a much smaller number of partial waves. The method is illustrated
for the ground states of hydrogenlike and lithiumlike boron, and the possibility
of high accuracy calculations on lower $Z$ hydrogenic ions is discussed.

\end{abstract}


\maketitle

\end{CJK*}

\section{introduction}\label{sec_intro}

The small $2s - 2p_{1/2}$ splitting in the spectrum of hydrogen
measured by Lamb and Retherford \cite{Lambexp} played a seminal role in
the development of Quantum Electrodynamics (QED) \cite{Schweber}.
The effect is generally referred to as the Lamb shift, and requires the evaluation of
two types of radiative corrections, vacuum polarization (VP) and the electron self energy (SE).
The first calculations exact to all orders in $Z\alpha$,
where $Z$ is the nuclear charge and $\alpha$ the fine structure constant,
%
%
\begin{equation}
\alpha \ \equiv \ \frac{e^2}{4\pi \epsilon_0 \hbar c} \ = \ 1 / 137.035\,999\,084(21),
\end{equation}
were carried out for VP by Wichmann and Kroll \cite{WK}.
Meanwhile, evaluations of SE assumed $Z\alpha$ was a small quantity
and could not be applied to cases when $Z$ was large. However, the possibility of
extending the work of Wichmann and Kroll to SE was realized, and not long after their work
the first all-order SE calculations were presented by
Desiderio and Johnson \cite{Desiderio} and by Mohr \cite{Mohr1}.
While both use partial wave expansions to represent the photon and electron propagators,
they differ in an important way in how the sum over all partial waves is carried out.

As will be seen in this paper, evaluation of the self-energy term always involves
multiple integrals over position $r$, a single integral over an energy $\omega$,
and an infinite sum over the partial wave $\ell$.
Mohr used a point-Coulomb potential,
and the fact that the electron propagator in this case can be expressed in terms of
Whittaker functions allowed him to sum the partial waves to convergence for any value of
coordinate and energy, even though for some values of the integrand, large sums were required.
This is the most accurate approach, and we will refer to it as the Mohr method in the following.

The work of Desiderio and Johnson (DJ), based on the method of
Brown {\it et al.}\ \cite{Brown}, was able to represent the electron propagator
in a general spherically symmetric potential, but because it used numerically generated
Green functions it was limited by how high a partial wave the numerical method could handle.
In this case the coordinate and energy integral was carried out for as many partial waves
as this limitation made possible, after which an extrapolation to higher values was made
for better convergence.

The work to be described here is a modification of a method we developed
in collaboration with Walter Johnson \cite{DJA}. It is based on Ref.\ \cite{Desiderio},
using a modification suggested by Snyderman \cite{Snyderman}
and implemented by Blundell and Snyderman \cite{Blundell}.
Because, as will be described below, an expansion of the propagator in terms of
an external potential is used, we call it the potential-expansion method as suggested
by Yerokhin, Pachucki and Shabaev \cite{Yerokhin1}.
In that paper, references to a number of different approaches can be found,
but the potential-expansion method closest to ours is given in \cite{Yerokhin2}.
%
%
As mentioned above, potential-expansion methods differ from the Mohr method in that
they carry out coordinate and energy integrations partial wave by partial wave,
afterward carrying out the final partial wave summation.
As in practice only a finite number of partial waves can be included, an extrapolation
of the partial wave expansion to infinity must be made, and there is a significant
numerical uncertainty associated with this procedure. In the following we will refer to
our original potential-expansion calculation as the DJA method. The purpose
of the present paper is to describe the DJB method, a modification
that improves the behavior of the partial wave series.

The most accurate calculations of the self energy have been obtained for point-Coulomb cases
using the Mohr method \cite{Mohr1999}. We illustrate the strikingly high accuracy the method
can attain with the self energy of the ground state of hydrogenlike boron which is given
by the dimensionless function $F_v(Z\alpha)$ as
\begin{equation}
  E_{SE}(v,Z) = \frac{\alpha}{\pi}\,\frac{(Z\alpha)^4}{n_v^3}\,F_v(Z\alpha)\,mc^2,
\label{eq_fza}
\end{equation}
with
\begin{equation}
  F_{1s}(5\alpha) = 6.251\,627\,078\,(1).
\label{eq_f1s}
\end{equation}
We will use this particular self energy in the following to illustrate details of
the DJA and DJB methods, and refer to it as the test case.
All self-energy results shown in this work will also be given in terms
of the dimensionless function $F_v(Z\alpha)$.

In Ref.~\cite{Yerokhin1}, it was pointed out that the potential-expansion method described
in Ref.~\cite{Yerokhin2} could be improved by using generalizations of identities used in
the Mohr method. These identities involve commuting (C) the potential (P) and
the free electron propagator (P), and we will refer to them as CPP identities.
They can be used to create approximations to terms of arbitrary order in
the potential expansion. These terms can be numerically evaluated very precisely,
but can also be expressed as partial wave expansions. In the latter form,
it is shown in Ref.~\cite{Yerokhin1} that when combined with the partial waves
computed in Ref.~\cite{Yerokhin2}, a much more tractable expansion results.
It is the purpose of this paper to describe how using a CPP identity allows us
to create a similarly improved partial wave expansion. Even with these improvements,
the potential-expansion method cannot reach the very high accuracy of Ref.~\cite{Mohr1999},
though in the treatment of the hydrogen isoelectronic sequence given in Ref.~\cite{Yerokhin3},
we note that the self-energy results presented are quite precise.

The plan of this paper is as follows. After the potential-expansion method is described
in Section \ref{sec_formalism}, Section \ref{sec_dja} describes the DJA method,
and the behavior of the first 30 partial waves for the test case is shown.
In Section \ref{sec_djb}, the DJB method is set up and the improvement of the
partial wave expansion shown for the test case.
In Section \ref{sec_application}, the DJB method is applied
to the $2s$ state of lithiumlike boron with a finite-nucleus model potential.
Finally, a discussion of how higher accuracy might be reached is given in the Conclusion.

\section{Formalism and subtraction schemes}\label{sec_formalism}

A central object in the self-energy calculation is the electron propagator, which satisfies the equation
\begin{equation}
  (z- H_x) G(z,{\bf x},{\bf y}) = \delta^3({\bf x - y})
\label{Gdef}
\end{equation}
with
\begin{equation}
  H_x = -i \hbar c\, \boldsymbol{\alpha} \cdot\! \boldsymbol{\nabla}_x+ mc^2\beta + V(x).
\end{equation}
We adopt the convention for any three-vector that $r \equiv |{\bf r}|$, so we are assuming our potential
to be spherically symmetric. In that case one can work with angular momentum eigenstates
characterized by the quantum numbers $\kappa$ and $\mu$. In the following, for simplicity we will assume
the potential to be that of a point nucleus of charge $Z$,
\begin{equation}
  V(r) = -\frac{Z e^2}{r}.
\end{equation}
Generalization to other potentials is straightforward. The Dirac equation
\begin{equation}
  H_x \psi_{v\kappa\mu}({\bf x})   = E_v \psi_{v\kappa\mu}({\bf x})
\end{equation}
has the solution
%
%
\begin{equation}
\psi_{v\kappa\mu}({\bf x}) =
   \left(
      \begin{array}{ll}
         \ g_v (x) \chi_{ \kappa \mu}(\hat{x}) \\
         i f_v (x) \chi_{-\kappa \mu}(\hat{x})
      \end{array}
   \right),
\label{psiv}
\end{equation}
with energy $E_v \equiv mc^2 \epsilon_v$.\\

A formally exact solution of Eq.\ (\ref{Gdef}) is obtained from a summation over all possible
$\kappa$ and $\mu$ values,
\begin{eqnarray}
G(z,{\bf x, y}) &=& \sum_{\kappa \mu} \Big[
  \theta(x-y) W_{\kappa \mu}(z,{\bf x}) U_{\kappa \mu}^\dagger(z,{\bf y})
\nonumber \\
&&+\ \theta(y-x) U_{\kappa \mu}(z,{\bf x}) W_{\kappa \mu}^\dagger(z,{\bf y}) \Big].
\end{eqnarray}
Here the spinors $U_{\kappa\mu}({z,\bf x})$ and $W_{\kappa\mu}(z,{\bf x})$ are of
the form of $\psi_{z\kappa\mu}({\bf x})$ in Eq.\ (\ref{psiv}), with the radial functions
being solutions to the Dirac equation regular at the origin and infinity, respectively.
Also, $\theta(t) = 0$ or 1 for $t <$ or $> 0$ is the step function.

The one-loop self energy of an electron state $v$
before regularization and renormalization is
%
\begin{align}
E^{(2)}_{SE} =& -4\pi i\alpha c^2 \!\int\! \frac{d^4k}{(2\pi)^4}
  \!\iint\! d^3 x \, d^3 y \,
  \frac{e^{i{\bf k \cdot (x-y)}}}{k_0^2 - {\bf k}^2 + i \delta} 
\nonumber \\ &\times
  \bar{\psi_v}({\bf x}) \gamma_{\nu} G(E_v - ck_0,{\bf x,y})
  \gamma_0 \gamma^{\nu} \psi_v({\bf y}).
\label{SEequation}
\end{align}
%
After renormalization, the finite remainder is the self-energy part
of the one-loop Lamb shift given by $E_{SE}(v,Z)$ in Eq.~(\ref{eq_fza}).
It has scaling factors $Z^4$ and $1/n_v^3$, and is of order $mc^2 \alpha^5$.

To evaluate $E^{(2)}_{SE}$ without expanding in $Z\alpha$ requires a treatment of the propagator
that allows regularization and removal of ultraviolet infinities, at the same time including the
finite parts of the calculation. To do this we expand the propagator around the free propagator
$F(z,{\bf x},{\bf y})$, which satisfies Eq.~(\ref{Gdef}) when $V(x)$ vanishes. The expansion is
\begin{widetext}
%
\begin{eqnarray}
G(z,{\bf x,y}) &=& F(z,{\bf x, y})
\nonumber \\
&& + \int d {\bf r}_1
  F(z,{\bf x, r}_1) V(r_1) F(z,{\bf r}_1, {\bf y})
\nonumber \\
&& +
  \iint d{\bf r}_1 d{\bf r}_2
  F(z,{\bf x, r}_1) V(r_1) F(z,{\bf r}_1, {\bf r}_2) V(r_2) F(z,{\bf r}_2, {\bf y})
\nonumber \\
&& +
  \iiint d{\bf r}_1 d{\bf r}_2 d{\bf  r}_3
  F(z,{\bf x, r}_1) V(r_1) F(z,{\bf r}_1, {\bf r}_2) V(r_2) F(z,{\bf r}_2, {\bf r}_3)
  V(r_3) F(z,{\bf r}_3, {\bf y})
\nonumber \\
&& + \ ...
\label{Gexpansion}
\end{eqnarray}
%
\end{widetext}

If we use the labeling scheme
\begin{equation}
  G(z,{\bf x},{\bf y}) \equiv \sum_{i=0}^{\infty} G_i(z,{\bf x}, {\bf y})
\end{equation}
where $i$ refers to the number of potentials $V(r)$ in the expansion terms
on the right-hand-side of Eq.\ (\ref{Gexpansion}),
the self energy can be similarly expanded,
\begin{equation}
  E^{(2)}_{SE} = \sum_{i=0}^{\infty} E_{iP},
\label{eq_emain}
\end{equation}
which defines the potential-expansion method.
All ultraviolet infinities are associated with the first two terms in the expansion, $E_{0P}$,
referred to as the zero-potential term, and $E_{1P}$, referred to as the one-potential term.
When these are separated from the complete sum, we define the result as the many-potential term,
\begin{equation}
  E_{MP} = \sum_{i=2}^{\infty} E_{iP}.
\label{eq_empsum}
\end{equation}
In the DJA method, $E_{MP}$ is evaluated in coordinate space,
and $E_{0P}$ and $E_{1P}$ in momentum space.
We now give a brief description of the calculation, with emphasis on $E_{0P}$,
modifications of which are used in the DJB method.

\section{DJA Method}\label{sec_dja}

To evaluate the zero- and one-potential terms we first define
the momentum space wave function,
\begin{equation}
  \psi_v({\bf p}) \equiv \int\! d^3 x \, e^{-i {\bf x}  \cdot {\bf p}/{\hbar}}\, \psi_v(x).
\end{equation}
The Dirac equation in momentum space is
\begin{equation}
  (\not p - mc)\, \psi_v({\bf p}) = -4\pi Z \alpha \!\int\!\! \frac{d{\bf p}_1}{(2\pi)^3}\,
  \frac{\gamma_0 \, \psi_v({\bf p}_1)}{|{\bf p}-{\bf p_1}|^2},
\label{pDirac}
\end{equation}
\\
where we have introduced  the 4-vector $p = (mc \epsilon, {\bf p})$. For the Dirac equation
$\epsilon = \epsilon_v$, but when $p$ is present in an electron propagator we leave $\epsilon$
as a variable that can be differentiated for later use when
we describe the DJB method.

We regulate the ultraviolet infinities in $E_{0P}$ and $E_{1P}$ by
changing $d^4 k \rightarrow d^n k$, with $n = 4 - \delta$. In dimensional regularization we
note that the self mass of a free electron at one-loop order is
\begin{equation}
  \delta m^{(2)} = m\, \frac{\alpha}{\pi} \left(\frac{3C}{\delta} + 2\right),
\end{equation}
with
\begin{equation}
  C = (4\pi)^{\delta/2} \,\Gamma(1+\delta/2).
\end{equation}

Using the representation of the free electron propagator
\begin{equation}
  F(z,{\bf x},{\bf y}) = \frac{1}{\hbar^3} \int\! \frac{d^3 p}{(2\pi)^3} \,
  \frac{e^{i {\bf p} \cdot ({\bf x - y})/\hbar}}
  {z \gamma_0 - c\, {\boldsymbol \gamma} \!\cdot\! {\bf p} -mc^2}\, \gamma_0,
\label{freedef}
\end{equation}
the zero-potential term can be shown to be
\begin{equation}
  E_{0P}(\epsilon) = -\frac{4 \pi i \alpha c}{\hbar^3} \int\! \frac{d{\bf p}}{(2\pi)^3} \,
  \bar{\psi}_v({\bf p}) X(p,\epsilon) \psi_v({\bf p}),
\end{equation}
with
\begin{align}
  X(p,\epsilon) \equiv& \int\! \frac{d^n k}{(2\pi)^n} \, \frac{1}{k^2} \, 
\nonumber \\ & \times
  \gamma_{\nu}
  \frac{1}{(mc \epsilon - k_0) \gamma_0 - \boldsymbol{\gamma} \!\cdot\! ({\bf p - k}) - mc}\,
  \gamma^{\nu}.
\end{align}
Standard manipulations give
\begin{widetext}
%
\begin{eqnarray}
X(p,\epsilon)
&=& \frac{1}{c} \int\! \frac{d^nk}{(2\pi)^n} \,
    \frac{\gamma_{\nu}(\not p - \not k + mc)\gamma^{\nu}}{k^2[(p-k)^2 - m^2c^2]}
\nonumber \\ 
&=& \frac{1}{c} \int_0^1\!dx \!\int\! \frac{d^n k}{(2\pi)^n} \,
    \frac{(2-n)(\not p - \not k) + n \cdot mc}{[(k-xp)^2+x(1-x)p^2 -x m^2c^2]^2}
\nonumber \\ 
&=& \frac{i C (mc)^{-\delta}}{8 \pi^2 c \delta} \!\int_0^1\!dx \,
    [\not p(1-x)(2-n) + n \cdot mc] \Delta^{-\delta/2},
\end{eqnarray}
%
%
with
\begin{equation}
  \Delta = x - x(1-x) \left[\epsilon^2 - {\bf p}^2/(mc)^2 \right].
\end{equation}
Expanding in $\delta$ and discarding terms of order $\delta$ and higher leads to
\begin{equation}
X(p,\epsilon) \ = \ \frac{i}{8\pi^2} \, m \left(\frac{3C}{\delta} + 2\right)
\,-\, \frac{i}{8\pi^2 c} \, (\not p - mc) \left(\frac{C}{\delta} + 1\right)
\,+\, \frac{i}{8\pi^2 c} \int_0^1 \! dx \big[\!\not p (1-x) - 2mc\,\big] \,
{\rm ln} \frac{\Delta}{x^2}.
\end{equation}
The zero-potential term is then
%
\begin{eqnarray}
E_{0P}(\epsilon) &=& \frac{\alpha}{2\pi}\,mc^2\left(\frac{3C}{\delta}+2\right)
  \frac{1}{\hbar^3} \int\! \frac{d{\bf p}}{(2\pi)^3} \,
  \bar{\psi}_v({\bf p}) \psi_v({\bf p})
\nonumber \\ 
&& - \frac{\alpha}{2\pi}\left(\frac{C}{\delta}+1\right) \frac{c}{\hbar^3}
  \int\! \frac{d{\bf p}}{(2\pi)^3} \bar{\psi}_v({\bf p}) (\not p -mc)
  \psi_v({\bf p})
\nonumber \\ 
 && + \ \frac{\alpha}{2\pi} \, \frac{c}{\hbar^3} \int\! \frac{d{\bf p}}{(2\pi)^3}
  \int_0^1 \!dx\, \bar{\psi}_v({\bf p}) \left[\not p(1-x) -2mc\right]
  \psi_v({\bf p}) \, {\rm ln} \frac{\Delta}{x^2}.
\label{E0Pe}
\end{eqnarray}
The first term in the right-hand-side is removed by mass renormalization.
\smallskip

Turning to $E_{1P}$, it is given by
\begin{equation}
E_{1P} = \frac{16\pi^2 icZ\alpha^2}{\hbar^3} \!\int\! \frac{d{\bf p}_2}{(2\pi)^3}
  \int\! \frac{d{\bf p}_1}{(2\pi)^3} \, \frac{1}{|{\bf p}_2 - {\bf p}_1|^2} \,
  \bar{\psi}_v({\bf p}_2) Y({\bf p}_2,{\bf p}_1) \psi_v({\bf p}_1),
\end{equation}
with
\begin{equation}
Y({\bf p}_2,{\bf p}_1) \equiv \int\! \frac{d^n k}{(2\pi)^n} \,
  \frac{\gamma_{\nu}(\not p_2 - \not k + mc) \gamma_0 (\not p_1 - \not k + mc)
  \gamma^{\nu}}{k^2\left[(k-p_2)^2-(mc)^2\right] \left[(k-p_1)^2 - (mc)^2\right]}.
\end{equation}
In the above, $p_1 = (mc \epsilon_v,\,{\bf p}_1)$
         and  $p_2 = (mc \epsilon_v,\,{\bf p}_2)$:
there is no need in this case to introduce $\epsilon$
and $\epsilon_v$ can be used directly.
A standard set of manipulations then leads to
\begin{eqnarray}
E_{1P} &=&
  \frac{\alpha}{2\pi} \left(\frac{C}{\delta} - \frac{1}{2}\right) \frac{c}{\hbar^3}
  \int\!\frac{d{\bf p}}{(2\pi)^3}\,\bar{\psi}({\bf p})(\not p - mc)\psi({\bf p})
\nonumber \\ 
&& + \ 2\, \frac{Z \alpha^2 c}{\hbar^3} \!
  \int\! \frac{d{\bf p}_2}{(2\pi)^3}
  \int\! \frac{d{\bf p}_1}{(2\pi)^3} \,
  \frac{{\bar\psi}({\bf p}_2) \gamma_0 \psi({\bf p}_1)} {|{\bf p}_2-{\bf p}_1|^2}
  \!\int_0^1\! \rho d \rho \!\int_0^1\! dx \, {\rm ln} \frac{\Delta_1}{\rho}
\nonumber \\ 
&& + \ \frac{Z \alpha^2 c}{\hbar^3}
  \int\! \frac{d{\bf p}_2}{(2\pi)^3}
  \int\! \frac{d{\bf p}_1}{(2\pi)^3}
  \int_0^1\! d \rho
  \int_0^1\! dx\, \frac{1}{\Delta_1} \,
  \frac{\bar{\psi}({\bf p}_2) N \psi({\bf p}_1)}{|{\bf p}_2 - {\bf p}_1|^2},
\end{eqnarray}
%
\end{widetext}
where the Dirac equation has been used in the first line and the
explicit form of $N$ can be found in Ref. \cite{DJA}. If we define
\begin{equation}
E_{01P} \equiv E_{0P} + E_{1P},
\end{equation}
after mass renormalization, we see it is ultraviolet finite. The
counter-terms present in the renormalization procedure that
would make the individual terms ultraviolet finite cancel because
of the Ward identity.

It is difficult to evaluate the finite part of $E_{1P}$ with high precision as it stands.
The solution used to improve the numerics is to employ the CPP identity introduced
by Mohr \cite{Mohr1}. The source of the numerical difficulties is the region where
$|\bf{p_1 - p_2}|$ is small. If we replace ${\bf p_2}$ with ${\bf p_1}$ everywhere
in the finite terms in $E_{1P}$ except the denominator and the wave function,
the Dirac equation, Eq.\ (\ref{pDirac}), can be used to carry out the $d{\bf p_2}$ integration,
resulting in a much simpler integral. By subtracting this term before using the Dirac equation,
the extra cancellation that results when ${|{\bf p_2}-{\bf p_1}|}^2$ is small
allows the integral to be evaluated with the accuracy needed.
(This procedure is carried out symmetrically, with ${\bf p_1}$ being replaced with
${\bf p_2}$ in a second subtraction term.) What we have just described is essentially
the procedure introduce by Mohr \cite{Mohr1} where the replacement of ${\bf p_2}$ with
${\bf p_1}$ in the propagator is the result of commuting the propagator through the potential.

The momentum space result for the test case from $0P$ and $1P$ is
\begin{equation}
  \ E_{01P} = -767.728\,102.
\label{0p1psum}
\end{equation}
The accuracy of the numerical integrations, which are done using the program CUHRE from
the Cuba package \cite{Cuba}, is such that all digits shown are significant. While the accuracy could be improved, there
would be no point in doing so because
the partial wave expansion involved in $E_{MP}$ leads to much larger numerical uncertainty.


We begin the coordinate space evaluation of $E_{MP}$ by carrying out the $d^3 k$ integration
in Eq.\ (\ref{SEequation}),
\begin{widetext}
\begin{equation}
E^{(2)}_{SE} = i \alpha \hbar c^2 \!\int\! \frac{dk_0}{2\pi}
  \! \iint \!  d^3 x\, d^3 y\,
  \frac{e^{i k_0 |{\bf x-y}|}} {|{\bf x}-{\bf y}|} \,
  \bar{\psi_v}({\bf x}) \gamma_{\nu} G(E_v - ck_0,{\bf x,y})
  \gamma_0 \gamma^{\nu} \psi_v({\bf y}).
\end{equation}
We define the order of the partial wave expansion $\ell$ by introducing the standard expansion
of the photon propagator,
\begin{equation}
\frac{e^{i k_0 |{\bf x-y}|}} {|{\bf x}-{\bf y}|} =
  \sum_{\ell=0}^{\infty} \, \sum_{m=-\ell}^{\ell} 4 \pi i k_0
  j_{\ell}(k_0 r) h_{\ell}(k_0 r') Y_{\ell m}(\hat{x}) Y^*_{\ell m}(\hat{y}),
\end{equation}
with $r={\rm min}(x,y)$ and $r'={\rm max}(x,y)$.
We now again use $\epsilon$, understood to be taken to
$\epsilon_v$ for the DJA method, and find
\begin{eqnarray}
E^{(2)}_{SE}(\epsilon) &=& i \alpha \hbar c^2
  \!\int\! \frac{dk_0}{2\pi} \sum_{\ell=0}^{\infty} \sum_{m=-\ell}^{\ell} 4\pi i k_0
  \! \iint \! d^3x\, d^3y\,
  j_{\ell}(k_0 r) h_{\ell}(k_0 r')
  Y_{\ell m}(\hat{x}) Y^*_{\ell m}(\hat{y})
\nonumber \\
&& \times \bigg[ \sum_{\kappa \mu} \theta(x-y) \bar{\psi_v}({\bf x}) \gamma_{\nu}
  U_{\kappa \mu}(\epsilon_v - k_0, \,{\bf x}) W_{\kappa \mu}^\dagger(\epsilon_v - k_0,\, {\bf y})
  \gamma_0 \gamma^{\nu} \psi_v({\bf y})
\nonumber \\
&& + \ \theta(y-x) \bar{\psi_v}({\bf x}) \gamma_{\nu}
  W_{\kappa \mu} (\epsilon_v - k_0, \,{\bf x}) {U^\dagger_{\kappa \mu}}(\epsilon_v - k_0, \,{\bf y})
  \gamma_0 \gamma^{\nu} \psi_v({\bf y}) \bigg] .
\label{eqe2se}
\end{eqnarray}
%
\end{widetext}
In this form one can analytically carry out the angle integrations along with the sum over $m$ and $\mu$.
The resulting Clebsch-Gordon coefficients then limit the sum over $\kappa$ for a given value of $\ell$,
and they are understood to be all included for any given partial wave.
Evaluation of the integrals over $k_0$, $x$, and $y$ can now be carried out if one has
the radial functions for the electron Green function, which are available analytically
in terms of Whittaker functions for the point-Coulomb case, or numerically for the
general case as in the present calculations.

%
\begin{table*}[t]
\caption{\label{tab1} DJA partial wave contributions to the self energy of the
$Z=5$ point-Coulomb $1s$ state. 
$E_{MP} = Main - E_{0P} - E_{1P}$. $Sum\_A$ is the cumulative partial-wave sum of $E_{MP}$.}
\begin{ruledtabular}
\begin{tabular}{cdddddd}
  \multicolumn{1}{c}{$\ell$}
& \multicolumn{1}{c}{$Main$}
& \multicolumn{1}{c}{$E_{0P}$}
& \multicolumn{1}{c}{$Main-E_{0P}$}
& \multicolumn{1}{c}{$E_{1P}$}
& \multicolumn{1}{c}{$E_{MP}$}
& \multicolumn{1}{c}{$Sum\_A$} \\
\colrule
 0 & 32953.2587\footnotemark[1]
                & 30259.7520 &2693.5067 &1937.5587 &755.9480 &-11.7801\footnotemark[2] \\
 1 & 34284.1223 & 33832.2160 & 451.9063 & 440.3507 & 11.5556 & -0.2245 \\
 2 & 34159.6544 & 33932.9018 & 226.7526 & 223.8197 &  2.9329 &  2.7084 \\
 3 & 33733.5632 & 33592.7097 & 140.8535 & 139.5826 &  1.2709 &  3.9792 \\
 4 & 33160.4627 & 33064.4550 &  96.0077 &  95.3232 &  0.6845 &  4.6637 \\
 5 & 32510.9396 & 32442.0604 &  68.8793 &  68.4636 &  0.4157 &  5.0794 \\
 6 & 31823.0807 & 31772.0957 &  50.9849 &  50.7121 &  0.2728 &  5.3522 \\
 7 & 31119.2099 & 31080.7241 &  38.4859 &  38.2968 &  0.1890 &  5.5412 \\
 8 & 30413.0504 & 30383.6582 &  29.3922 &  29.2557 &  0.1365 &  5.6777 \\
 9 & 29713.2633 & 29690.6918 &  22.5715 &  22.4698 &  0.1017 &  5.7794 \\
10 & 29025.3708 & 29008.0375 &  17.3334 &  17.2556 &  0.0777 &  5.8571 \\
11 & 28352.8744 & 28339.6399 &  13.2345 &  13.1738 &  0.0607 &  5.9178 \\
12 & 27697.9380 & 27687.9599 &   9.9781 &   9.9298 &  0.0482 &  5.9661 \\
13 & 27061.8230 & 27054.4646 &   7.3584 &   7.3194 &  0.0389 &  6.0050 \\
14 & 26445.1730 & 26439.9441 &   5.2289 &   5.1971 &  0.0318 &  6.0368 \\
15 & 25848.2057 & 25844.7229 &   3.4828 &   3.4565 &  0.0263 &  6.0631 \\
16 & 25270.8445 & 25268.8038 &   2.0407 &   2.0187 &  0.0220 &  6.0851 \\
17 & 24712.8100 & 24711.9676 &   0.8424 &   0.8238 &  0.0185 &  6.1037 \\
18 & 24173.6846 & 24173.8431 &  -0.1585 &  -0.1743 &  0.0158 &  6.1194 \\
19 & 23652.9589 & 23653.9569 &  -0.9980 &  -1.0114 &  0.0135 &  6.1329 \\
20 & 23150.0646 & 23151.7690 &  -1.7043 &  -1.7160 &  0.0116 &  6.1446 \\
21 & 22664.3984 & 22666.6988 &  -2.3003 &  -2.3104 &  0.0101 &  6.1547 \\
22 & 22195.3394 & 22198.1435 &  -2.8041 &  -2.8129 &  0.0088 &  6.1635 \\
23 & 21742.2613 & 21745.4918 &  -3.2305 &  -3.2382 &  0.0077 &  6.1712 \\
24 & 21304.5417 & 21308.1331 &  -3.5914 &  -3.5982 &  0.0068 &  6.1780 \\
25 & 20881.5681 & 20885.4650 &  -3.8969 &  -3.9029 &  0.0060 &  6.1840 \\
26 & 20472.7424 & 20476.8976 &  -4.1552 &  -4.1606 &  0.0053 &  6.1893 \\
27 & 20077.4842 & 20081.8573 &  -4.3732 &  -4.3779 &  0.0048 &  6.1941 \\
28 & 19695.2324 & 19699.7889 &  -4.5565 &  -4.5607 &  0.0043 &  6.1983 \\
29 & 19325.4469 & 19330.1569 &  -4.7100 &  -4.7138 &  0.0038 &  6.2022 \\
30 & 18967.6086 & 18972.4465 &  -4.8379 &  -4.8413 &  0.0034 &  6.2056 \\
\colrule
\multicolumn{5}{l}{High-$\ell$ correction from $1/\ell^3$ fit} & 0.0500 & 6.2557 \\
\multicolumn{5}{l}{High-$\ell$ correction $\Delta\ell^{\,-3}_{2,5}$
                   from Eq.~(\ref{eq_fit})} & 0.0465 & 6.2521 \\
\colrule
\multicolumn{6}{l}{Ref.~\cite{Mohr1999}}  &  6.2516 \\
\end{tabular}
\end{ruledtabular}
\footnotetext[1]{Include contribution from the pole term in Eq.\ (\ref{eqpole}).}
\footnotetext[2]{Include momentum-space contribution from $E_{01P}$ in Eq.\ (\ref{0p1psum}).}
\end{table*}

In the DJA method,
integrations over $\omega = c k_0$ are carried out for each
partial wave $\ell$, and the resulting  partial wave series is summed
to give the final results. $E^{(2)}_{SE}$ thus calculated will be referred to as the {\it Main} term.
To form the ultraviolet convergent many-potential term $E_{MP} = E^{(2)}_{SE} - E_{0P} - E_{1P}$,
we begin by subtracting the zero-potential term $E_{0P}$ from the {\it Main} term.
Computationally, $E_{0P}$ in coordinate space is the same as $E^{(2)}_{SE}$
with the bound-electron Green function $G(z,{\bf x},{\bf y})$
replaced by the free-electron Green function $F(z,{\bf x},{\bf y})$
which can be generated analytically or numerically.
Partial waves of the {\it Main} and $E_{0P}$ terms up to $\ell=30$ are shown
in the second and third columns of Table~\ref{tab1} for the 
test case and their difference is shown in the fourth column. It is clear that there are
substantial cancellations between {\it Main} and $E_{0P}$, but $Main-E_{0P}$ is
a partial wave expansion that does not converge, and the gradual falloff with $\ell$
eventually goes as $1/\ell$, which corresponds to a logarithmic divergence.

We note 
that in evaluating $E_{SE}^{(2)}$, 
a Wick rotation, $\omega \rightarrow i \omega$,
is carried out, and a deformation of the contour
to avoid bound-state poles gives rise to the ``Pole terms''. Details can be found
in Ref.~\cite{DJA}. Pole terms do not involve electron Green functions and
can be calculated very accurately. For the $E_{1s}(5\alpha)$ test case
considered here, there is only one $1s$ pole term given by
\begin{equation}
  E_{1s}({\rm pole}) = 20\,210.432\,546.
\label{eqpole}
\end{equation}
This term is combined with the $\ell=0$ partial wave of the {\it Main} term
in Table~\ref{tab1},
as this is the only partial wave affected by the $1s$ pole from
symmetry and energy considerations.

To finally form the ultraviolet finite many potential term, we need to compute
\begin{widetext}
%
\begin{eqnarray}
E_{1P} &=& i \alpha \hbar c^2 \!\!\int\!\! \frac{dk_0}{2\pi}
  \! \iint \! d^3\!x\, d^3\!y\,
  \frac{e^{i k_0 |{\bf x-y}| }} {|{\bf x-y}|} \,
  \bar{\psi_v}({\bf x}) \gamma_{\nu} G_2(E_v - ck_0,{\bf x,y})
  \gamma_0 \gamma^{\nu} \psi_v({\bf y})
\nonumber \\ [4pt]
&=& i \alpha \hbar c^2 \!\!\int\!\! \frac{dk_0}{2\pi}
  \!\! \iiint \!\! d^3\!x\, d^3\!w\, d^3\!y \,
  \frac{e^{i k_0 |{\bf x-y}| }} {|{\bf x-y}|} \,
  \bar{\psi_v}({\bf x}) \gamma_{\nu} F(E_v - ck_0,{\bf x,w})
\gamma_0 \frac{\hbar c Z \alpha}{w}
  F(E_v-ck_0,{\bf w,y}) \gamma_0 \gamma^{\nu} \psi_v({\bf y}).
\end{eqnarray}
%
\end{widetext}
The ordering of the magnitudes $x$, $w$, and $y$ determines which spherical Bessel functions
must be used, and requires the evaluation of three different integrals. However,
only one more integration variable is present compared to the zero potential term,
and no numerical difficulties arise. The result is presented in the
fifth column of Table~\ref{tab1}.
Subtractions of $E_{1P}$ from $Main - E_{0P}$ give the $E_{MP}$ term listed
in the sixth column. Once again, there are substantial cancellations, but the resulting
partial wave series of $E_{MP}$ now converges as $1/\ell^3$.

In the seventh column of Table \ref{tab1}, the cumulative partial-wave sum of $E_{MP}$
are shown as $Sum\_A$.
By adding $E_{01P}$ in Eq.~(\ref{0p1psum}) to the $\ell = 0$ term,
$Sum\_A$ should converge to the final self-energy result.
Indeed, $Sum\_A(\ell)$ can be seen to approach the high-precision results
of 6.2516$\ldots$ from Ref.~\cite{Mohr1999}, 
with $Sum\_A(30) = 6.2056$ converged to the
first decimal point for an accuracy of 0.74\%.
By extrapolating the partial wave series with an
$1/\ell^3$ fit, the high-$\ell$ contribution from $\ell = 31 - \infty$
of 0.0500 can be added to $Sum\_A(30)$ for a result of 6.2557 shown in the third
row from the bottom in Table~\ref{tab1}.
This improves the convergence by one more decimal point and the accuracy to 0.06\%.
High-$\ell$ corrections have also been calculated with an accelerated-convergence method
based on a $k$-point least-square, rational polynomial fit of the form
\begin{equation}
  f^{\,-n}_{m,k}(\ell) \approx 1/[\ell^n(a_0 + a_1/\ell + \cdots + a_m/\ell^m)],
\label{eq_fit}
\end{equation}
where the number of least-square points $k$ must be greater than
the order of the rational polynomial $m$.
In fact, the $1/\ell^3$ fit is a special case with $n=3$, $m=0$ and $k=1$.
As shown in the second row from the bottom of Table~\ref{tab1},
the high-$\ell$ correction $\Delta\ell^{\,-3}_{2,5}$ of 0.0465 does
accelerate the convergence and further
improves the self-energy result
by another decimal point to 6.2521 for an accuracy of 0.01\%.

While the difference between the two high-$\ell$ extrapolation results reflects the
intrinsic uncertainty of these corrections, their contributions can be greatly
reduced by extending the calculation to include more partial waves.
For higher-$Z$ ions than the $Z=5$ test case, that is usually not necessary
as partial wave series tend to converge much faster.
For lower-$Z$ ions, however, partial wave series
converge much slower, and unlike the Mohr method that utilizes analytic functions
extensively, the numerical approach of the DJA method limits the number of partial waves
that can be accurately calculated.
A different approach with faster partial wave convergence is needed.
For that, we turn to the new DJB method which is based on a variation of the method
in Ref.~\cite{Yerokhin1}.

\section{DJB Method}\label{sec_djb}

The next logical step in the potential-expansion method would appear to be the evaluation
of $E_{2P}$, given by
\begin{widetext}
%
\begin{eqnarray}
E_{2P} & = & -4 \pi i \alpha c^2 \!\int\!
  \frac{d^4k}{(2\pi)^4}
  \! \iint \! d^3 x \, d^3 y \,
  \frac{e^{i {\bf k} \cdot ({\bf x-y})}} {k_0^2- {\bf k}^2 + i \delta} \,
  \bar{\psi_v}({\bf x}) \gamma_{\nu}
\nonumber \\ [4pt]
&& \times
  \iint \! d{\bf r}_1 d{\bf r}_2 \,
  F(E_v - ck_0,{\bf x,r}_1) V(r_1)
  F(E_v- ck_0,{\bf r}_1,{\bf r}_2) V(r_2) F(E_v - ck_0,{\bf r}_2,{\bf y})
 \gamma^{\nu} \psi_v({\bf y}).
\end{eqnarray}
However, after transforming to momentum space and evaluating the $d^4 k$ integral
with Feynman parameters, one has a multidimensional integral of nominal dimension $9$.
Evaluating such an integral to high precision would be an extremely challenging proposition
even with the subtractions described for $E_{1P}$.
We consider instead an approximation $\tilde{E}_{2P}$,
\begin{eqnarray}
\tilde{E}_{2P} & \equiv & -4 \pi i \alpha c^2 \!\int\!
  \frac{d^4k} {(2\pi)^4}
      \! \iint \!  d^3 x \, d^3 y \,
      \frac{e^{i {\bf k} \cdot ({\bf x-y})}} {k_0^2- {\bf k}^2 + i \delta} \,
  \bar{\psi_v}({\bf x}) \gamma_{\nu}
\nonumber \\ [4pt]
&& \times \
  \! \iint \! d {\bf r_1} d {\bf r_2} \,
  F(E_v - c k_0,{\bf x},{\bf r_1}) V(x) F(E_v - c k_0,{\bf r_1},{\bf r_2}) V(y)
  F(E_v - ck_0,{\bf r_2},{\bf y})
\gamma^{\nu} \psi_v({\bf y}).
\end{eqnarray}
%
\end{widetext}
Because the free electron propagator $F(z,{\bf x},{\bf y})$ emphasizes the region
${\bf x} = {\bf y}$, the replacement of $V(r_1)$ with $V(x)$ and $V(r_2)$ with $V(y)$
in $\tilde{E}_{2P}$ can be expected to capture a dominant part of the integral.
The replacement corresponds to a CPP method, with $V(r_1)$ commuted to the left
and $V(r_2)$ commuted to the right.

The DJB method involves replacing the MP term in the DJA method with
\begin{equation}
E_{MP} \,=\, (E_{MP}-{\tilde{E}}_{2P}) + \tilde{E}_{2P}
\,\equiv\,  {\tilde{E}}_{MP} + \tilde{E}_{2P}.
\label{eq_Emp}
\end{equation}
The relative simplicity of $\tilde{E}_{2P}$ comes from the identity
\begin{widetext}
\begin{equation}
  \iint \! d^3u\, d^3w\,
  F(z,{\bf x, u}) F(z,{\bf u, w}) F(z,{\bf w, y})
  = \frac{1}{2} \, \frac{d^2}{dz^2} F(z,{\bf x, y}).
\end{equation}
This allows the ${\bf r_1}$ and ${\bf r_2}$ integrations in $\bar{E}_{2P}$
to be carried out, and we have
\begin{equation}
\tilde{E}_{2P}  = -2 \pi i \alpha c^2 \frac{d^2}{dE_v^2}
  \int\! \frac{d^4k}{(2\pi)^4}
  \! \iint \! d^3 x \, d^3 y \, V(x) V(y) \,
  \frac{e^{i {\bf k} \cdot ({\bf x-y}) }}
  {k_0^2- {\bf k}^2 + i \delta}
\bar{\psi_v}({\bf x}) \gamma_{\nu} F(E_v-ck_0,{\bf x, y})
  \gamma_0 \gamma^{\nu} \psi_v({\bf y}).
\label{e3v}
\end{equation}
%
\end{widetext}
In coordinate space form, this is to be subtracted from $E_{MP}$,
and to compensate we need to add it back in momentum space form.
\smallskip

To do this, we work with Eq.\ (\ref{E0Pe}), which we treated as a function of $\epsilon$.
$\tilde{E}_{2P}$ involves differentiating with respect to $\epsilon$ twice,
after which case one can take $\epsilon \rightarrow \epsilon_v$. We start by noting
\begin{align}
\frac{d^{2\!} X(p)}{d\epsilon_v^2} =&\ \frac{i}{4 \pi^2 m^2 c^5}
  \int_0^1\! dx \, x(1-x)
\nonumber \\ & \times
\left[\frac{\epsilon_v N_0-N_1}{\Delta} +
  \frac{2 D_B(\epsilon_v N_0 + N_1)}{\Delta^2}\right],
\end{align}
where $N_0 = -\gamma_0(1-x)/c$ and $N_1 = 2mc + \boldsymbol{\gamma} \cdot {\bf p}\,(1-x)$.
If we define the momentum space function
\begin{equation}
\psi_{v_1}({\bf p}) \equiv \int\! d^3 x \, e^{-i {\bf x} \cdot {\bf p}/\hbar} \,
  \frac{Z(x)}{x} \, \psi_v(x)
\end{equation}
one has
%
\begin{align}
\tilde{E}_{2P} =&\ \frac{\alpha}{3\pi m^2 c^3 \hbar^3}
  \!\int\! \frac{d{\bf p}}{(2\pi)^3}
  \!\int_0^1\! \!dx \, x(1-x)
  \bar{\psi}_{v_1}({\bf p})
\nonumber \\ & \times
  \!\left[\frac{\epsilon_v N_0-N_1}{\Delta} +
  \frac{2 D_B(\epsilon_v N_0 + N_1)}{\Delta^2}\right]\! \psi_{v_1}({\bf p}),
\end{align}
%
which can be easily evaluated with high accuracy. Its value in momentum space
for the test case is given by
\begin{equation}
  \tilde{E}_{2P} = 365.613\,427.
\label{2psum}
\end{equation}

Turning to the coordinate space part of the calculation,
we note that the double derivative with respect to $E_v$ can be carried out
using the recursion relations for spherical Bessel functions.
While this results in a somewhat complicated integrand, the numerical integral
is of the same form as used for the other parts of the coordinate space calculation,
and the results are of the same accuracy.
We list the partial waves for the test case up to
$\ell=30$ in the third column of Table~\ref{tab2}.

A check on the calculation can now be made by
comparing the partial wave expansion of $\tilde{E}_{2P}$ with the
momentum space form, which can be evaluated with high precision.
From Table~\ref{tab2}, the partial wave sum of $\tilde{E}_{2P}$ up to $\ell=30$
is 365.5675, which agrees with the momentum space result in Eq.~(\ref{2psum}) to 0.01\%.
While this check reflects the accuracy possible
for the partial wave expansion, that accuracy is still limited
for the same reason DJA is limited, the relatively slow convergence
of the partial wave expansion. However, the partial wave expansion of DJB has two features
that make the method much more accurate.
The first is that the cancellation
with $E_{MP}$, shown in the sixth column of Table~\ref{tab1} and
re-shown in the second column of Table~\ref{tab2},
makes the higher partial waves smaller by over
two orders of magnitude as shown in the fourth column of
Table~\ref{tab2}. Indeed, the cumulative sum of $\tilde{E}_{MP} = E_{MP} - \tilde{E}_{2P}$,
shown as $Sum\_B$ in the fifth column,
can be seen to converge readily to 6.2515 at $\ell=30$ instead of $Sum\_A$'s 6.0256
in Table~\ref{tab1}.
The second feature is that the convergence of $\tilde{E}_{MP}$ is more rapid
at $1/\ell^4$. Using a range of extrapolation methods as done with the DJA method,
we find that at $\ell = 30$, they all give consistent high-$\ell$
corrections at $\sim$0.0001, improving the present self-energy result to 6.2516,
same as the high-precision results of Ref.~\cite{Mohr1999} down to the 4th decimal point
as seen in the last two rows in Table~\ref{tab2}.
For higher-$Z$ ions than the present test case of $Z=5$, $10 - 20$ partial waves would
likely be sufficient for the DJB method, and high-$\ell$ extrapolations may not
even be necessary except for accuracy checks. DJB is a marked improvement over DJA.

%
\begin{table}[t]
\caption{\label{tab2} DJB partial wave contributions to the self energy of the
$Z=5$ point-Coulomb $1s$ state.
$\tilde{E}_{MP} = E_{MP}-\tilde{E}_{2P}$.
$Sum\_B$ is the cumulative partial-wave sum of $\tilde{E}_{MP}$.}
\begin{ruledtabular}
\begin{tabular}{cdddd}
  \multicolumn{1}{c}{$\ell$}
& \multicolumn{1}{c}{$E_{MP}$}
& \multicolumn{1}{c}{$\tilde{E}_{2P}$}
& \multicolumn{1}{c}{$\tilde{E}_{MP}$}
& \multicolumn{1}{c}{$Sum\_B$} \\
\colrule
 0  & 755.9480  & 349.1413  & 406.8067  &  4.6921\footnotemark[1] \\
 1  &  11.5556  &  10.1811  &   1.3744  &  6.0665 \\
 2  &   2.9329  &   2.8131  &   0.1198  &  6.1863 \\
 3  &   1.2709  &   1.2374  &   0.0335  &  6.2198 \\
 4  &   0.6845  &   0.6708  &   0.0136  &  6.2334 \\
 5  &   0.4157  &   0.4090  &   0.0067  &  6.2401 \\
 6  &   0.2728  &   0.2690  &   0.0038  &  6.2439 \\
 7  &   0.1890  &   0.1868  &   0.0023  &  6.2462 \\
 8  &   0.1365  &   0.1350  &   0.0015  &  6.2476 \\
 9  &   0.1017  &   0.1007  &   0.0010  &  6.2486 \\
10  &   0.0777  &   0.0771  &   0.0007  &  6.2493 \\
11  &   0.0607  &   0.0602  &   0.0005  &  6.2498 \\
12  &   0.0482  &   0.0479  &   0.0004  &  6.2502 \\
13  &   0.0389  &   0.0387  &   0.0003  &  6.2505 \\
14  &   0.0318  &   0.0316  &   0.0002  &  6.2507 \\
15  &   0.0263  &   0.0262  &   0.0002  &  6.2508 \\
16  &   0.0220  &   0.0219  &   0.0001  &  6.2510 \\
17  &   0.0185  &   0.0184  &   0.0001  &  6.2511 \\
18  &   0.0158  &   0.0157  &   0.0001  &  6.2512 \\
19  &   0.0135  &   0.0134  &   0.0001  &  6.2512 \\
20  &   0.0116  &   0.0116  &   0.0001  &  6.2513 \\
21  &   0.0101  &   0.0100  &   0.00005 &  6.2513 \\
22  &   0.0088  &   0.0088  &   0.00004 &  6.2514 \\
23  &   0.0077  &   0.0077  &   0.00003 &  6.2514 \\
24  &   0.0068  &   0.0068  &   0.00003 &  6.2514 \\
25  &   0.0060  &   0.0060  &   0.00002 &  6.2515 \\
26  &   0.0053  &   0.0053  &   0.00002 &  6.2515 \\
27  &   0.0048  &   0.0047  &   0.00002 &  6.2515 \\
28  &   0.0043  &   0.0042  &   0.00002 &  6.2515 \\
29  &   0.0038  &   0.0038  &   0.00001 &  6.2515 \\
30  &   0.0034  &   0.0034  &   0.00001 &  6.2515 \\
\colrule
\multicolumn{3}{l}{High-$\ell$ correction $\Delta\ell^{\,-4}_{2,5}$}
                   & 0.00010 & 6.2516 \\
\colrule
\multicolumn{3}{l}{Ref.~\cite{Mohr1999}}    &         &  6.2516 \\
\end{tabular}
\end{ruledtabular}
\footnotetext[1]{Include momentum-space contributions from $E_{01P}$ in Eq.~(\ref{0p1psum})
and $\tilde{E}_{2P}$ in Eq.~(\ref{2psum}).}
\end{table}

\section{Applications}\label{sec_application}

Now that we have shown the details of the DJB method for the test case,
it is clear that the present DJB result of $E_{1s}(5\alpha)=6.2516(1)$,
with an uncertainty of 1 in the last digit,
cannot match the accuracy of 6.251\,627\,078(1)
in Ref.~\cite{Mohr1999}. Nevertheless, the present result is still accurate
to 5 significant figures, more than enough for most applications.
More importantly, the present approach is not limited to point-Coulomb cases,
as bound state wave functions and electron Green functions are solved
numerically instead of derived analytically. Thus, DJB method has a wide range
of applications and can be used to calculate, for example, electron screening
and finite-nuclear size corrections to electron self energies.

Choosing the $2s$ state of Li-like boron as an example, we start by
using a Kohn-Sham potential for the $1s^2 2s$ ground state to account for
the screening effect. Finite-nuclear size potential is modeled by
a Fermi charge distribution with parameters $c=1.8104$~fm and $t=2.3$~fm.
%
%
%
Partial wave results up to $\ell=20$ are shown in Table~\ref{tab3}.
Specifically, $E_{MP}$ in column~2 and $Sum\_A$ in column~3
correspond to DJA results with one-potential expansions,
while $\tilde{E}_{2P}$ in column~4, $E_{MP}-\tilde{E}_{2P}$ in column~5 and
$Sum\_B$ in columns~6 are DJB results with the additional two-potential expansions.
High-$\ell$ corrections $\Delta\ell^{\,-4}_{2,5}$ to $Sum\_B$
from least-square rational polynomial fits of the form given in Eq.~(\ref{eq_fit})
are shown for $\ell \ge 5$ in column~7
and $Total\_B=Sum\_B+\Delta\ell_{2,5}^{\,-4}$ are shown in column~8.
As in Tables~\ref{tab1} and \ref{tab2}, the $E_{MP}$, $Sum\_A$ and $Sum\_B$ terms
have the pole and momentum-space terms included in the $\ell=0$ partial waves
so that the cumulative sums of $Sum\_A$, $Sum\_B$ and $Total\_B$ will converge
to the self energy $E_{2s}(5\alpha)$.

At the DJA level, it can be seen that $E_{MP}(20)$ only goes down to 0.0132
and $Sum\_A(20)$, at 2.7855, is far from convergence.
With DJB, however, $E_{MP}-\tilde{E}_{2P}$ is already down to 0.0000\,3 at $\ell=20$,
and $Sum\_B(20)$, at 2.9567, is nearly converged to the last digit.
When \mbox{high-$\ell$} corrections $\Delta\ell^{\,-4}_{2,5}$ are added,
$Total\_B$ actually converges to 2.9569 with only 8 partial waves
even though high-$\ell$ correction is still rather large at 0.0025.
While $\Delta\ell^{\,-4}_{2,5}$ continues to drop by an order-of-magnitude
to 0.0002 at $\ell=20$, $Total\_B$ remains constant to the fourth decimal point.
This is a good check on the accuracy of the final result and affirms
the use of high-$\ell$ extrapolation methods to accelerate the
partial wave convergence.
Comparing to the test case, it is clear that the DJB method converges much faster
with non-Coulomb potentials even for higher-$n$ ($2s$ vs.\ $1s$) states.
There is also no doubt that DJB is an important improvement over DJA even though
the later can give accurate enough results in most cases with larger partial
wave expansions.

%
\begin{table*}[t]
\caption{\label{tab3}
DJA and DJB partial wave contributions to the self energy of the
$2s$ state of Li-like boron ($Z=5$) in a finite-nucleus, Kohn-Sham potential.
$\tilde{E}_{MP}=E_{MP}-\tilde{E}_{2P}$.
$Sum\_A$ and $Sum\_B$ are cumulative partial-wave sums of
$E_{MP}$ and $\tilde{E}_{MP}$, respectively.
$\Delta\ell^{\,-4}_{2,5}$ are high-$\ell$ corrections of $Sum\_B$
by fitting $\tilde{E}_{MP}$ with Eq.~(\ref{eq_fit}).
$Total\_B = Sum\_B + \Delta\ell^{-4}_{2,5}$.
}
\begin{ruledtabular}
\begin{tabular}{cddddddd}
  \multicolumn{1}{c}{$\ell$}
& \multicolumn{1}{c}{\quad$E_{MP}$}
& \multicolumn{1}{c}{\quad$Sum\_A$}
& \multicolumn{1}{c}{\quad$\tilde{E}_{2P}$}
& \multicolumn{1}{c}{\quad$\tilde{E}_{MP}$}
& \multicolumn{1}{c}{~\quad$Sum\_B$}
& \multicolumn{1}{c}{$\Delta\ell^{\,-4}_{2,5}$}
& \multicolumn{1}{c}{$Total\_B$} \\
\colrule
 0 & 577.4753\footnotemark[1]
              & -8.2795\footnotemark[2]
                        & 298.9027 & 278.5726  & 2.2123\footnotemark[3] \\
 1 &   6.9704 & -1.3091 &   6.3045 &   0.6659  & 2.8782 \\
 2 &   1.7532 &  0.4441 &   1.7022 &   0.0509  & 2.9291 \\
 3 &   0.7848 &  1.2289 &   0.7707 &   0.0141  & 2.9432 \\
 4 &   0.4390 &  1.6679 &   0.4333 &   0.0058  & 2.9490 \\
 5 &   0.2773 &  1.9452 &   0.2744 &   0.0029  & 2.9518 & 0.0052  & 2.9570 \\
 6 &   0.1893 &  2.1345 &   0.1877 &   0.0016  & 2.9535 & 0.0035  & 2.9570 \\
 7 &   0.1365 &  2.2710 &   0.1355 &   0.0010  & 2.9544 & 0.0025  & 2.9569 \\
 8 &   0.1024 &  2.3734 &   0.1018 &   0.0006  & 2.9551 & 0.0018  & 2.9569 \\
 9 &   0.0793 &  2.4527 &   0.0789 &   0.0004  & 2.9555 & 0.0014  & 2.9569 \\
10 &   0.0630 &  2.5157 &   0.0627 &   0.0003  & 2.9558 & 0.0011  & 2.9569 \\
11 &   0.0511 &  2.5668 &   0.0508 &   0.0002  & 2.9560 & 0.0009  & 2.9569 \\
12 &   0.0421 &  2.6089 &   0.0419 &   0.0002  & 2.9562 & 0.0007  & 2.9569 \\
13 &   0.0352 &  2.6441 &   0.0351 &   0.0001  & 2.9563 & 0.0006  & 2.9569 \\
14 &   0.0298 &  2.6739 &   0.0297 &   0.0001  & 2.9564 & 0.0005  & 2.9569 \\
15 &   0.0255 &  2.6994 &   0.0254 &   0.0001  & 2.9565 & 0.0004  & 2.9569 \\
16 &   0.0220 &  2.7214 &   0.0220 &   0.0001  & 2.9565 & 0.0003  & 2.9569 \\
17 &   0.0192 &  2.7406 &   0.0191 &   0.00005 & 2.9566 & 0.0003  & 2.9569 \\
18 &   0.0168 &  2.7575 &   0.0168 &   0.00004 & 2.9566 & 0.0003  & 2.9569 \\
19 &   0.0149 &  2.7723 &   0.0148 &   0.00003 & 2.9567 & 0.0002  & 2.9569 \\
20 &   0.0132 &  2.7855 &   0.0132 &   0.00003 & 2.9567 & 0.0002  & 2.9569 \\
\end{tabular}
\end{ruledtabular}
\footnotetext[1]{Include the pole contribution of 34\,654.400\,87.}
\footnotetext[2]{Include the $E_{01P}$ contribution of $-585.754\,79$.}
\footnotetext[3]{Include the $E_{01P} + \tilde{E}_{2P}$ contribution of $-276.360\,31$.}
\end{table*}

\section{Conclusion}
We have deliberately used only
a modest number of partial waves in this paper.
This is because we wish to emphasize that relatively simple calculations can allow
quite accurate self energies to be computed. However, one application that requires
extremely high accuracy is the self energy of hydrogen. As with our test case,
it has been evaluated with extremely high accuracy in Ref.~\cite{Mohr1999}. 
Because that accuracy is needed in the treatment of the finite size of the proton,
a check using different methods would be useful.

One of the advantages already present in the DJA method is the fact that
the numerical methods used allow one to go up to values of $\ell\approx 60$,
though extreme care and very fine radial grids are needed.
In fact, it is possible to control the high-$\ell$ extrapolation
so well that the DJB method is usually not qualitatively more accurate,
but it works just as well as DJA with fewer partial waves
and can deal with low $Z$ better than DJA. DJB thus supersedes DJA as
a general approach to self-energy calculations.
However, calculations at $Z=1$ of radiative corrections are particularly challenging
even for DJB.
Sophisticated summation schemes were required in the framework of the Mohr method in Ref.
\cite{Mohr1999} to reach the very high accuracy results presented there.
To reach similar accuracy with potential-expansion methods, many numerical issues
would have to be addressed. Techniques that are more than adequate for calculations
demanding part per million accuracy may fail at higher levels.

We are at present working on evaluating the self energy of
hydrogen and hydrogenic ions with low $Z$. Because of the
numerical problems that may be present that we have not
detected, we are also looking into the use of different gauges.
There are advantages to the use of both Coulomb gauge and Yennie
gauge that are known to help with the infrared behavior of
radiative corrections. While very accurate calculations have
in fact already been carried out in Feynman gauge, getting
the same result using another gauge would clearly be a
check on the calculation, and getting different answers
could uncover numerical problems that had been missed.
However, we conclude by emphasizing the utility and
relative ease of using the DJB
method described in this paper for those interested in
evaluating the self energy part of the Lamb shift.

\acknowledgments

The work of KTC was performed under the auspices of the U.S. Department of Energy
by Lawrence Livermore National Laboratory under Contract DE-AC52-07NA27344.
We would like to thank Walter Johnson,
Peter Mohr, and Vladimir Yerokhin for useful conversations.



\end{document}